\documentclass{aa}
\input epsf

\begin{document}
\sloppypar

   \thesaurus{06    
              (08.02.3;  
               08.06.1;  
               08.14.1;  
               08.09.2 XTE J1806--246;  
               13.25.5;  
               19.63.1)} 

   \title{Discovery of 0.08 Hz QPO in the power spectrum of black hole
   candidate XTE~J1118+480}

   \author{M. Revnivtsev \inst{1,2}, R. Sunyaev\inst{2,1},
   K. Borozdin \inst{3,1}}

   \offprints{revnivtsev@hea.iki.rssi.ru}

   \institute{Space Research Institute, Russian Academy of Sciences,
              Profsoyuznaya 84/32, 117810 Moscow, Russia 
        \and
              Max-Planck-Institute f\"ur Astrophysik,
              Karl-Schwarzschild-Str. 1, D-85740 Garching bei M\"unchen,
              Germany,
	\and
		NIS-2, Los Alamos National Laboratory, NM 87545, USA
            }
  \date{}

        \authorrunning{Revnivtsev, Sunyaev \& Borozdin}
        \titlerunning{QPO in XTE~J1118+480} 
        
   \maketitle

\begin{abstract}
We found a strong QPO feature at $0.085\pm0.002$ Hz in 
the power spectrum of X-ray transient XTE J1118+480.
The QPO was detected in PCA/RXTE data with an amplitude 
close to 10\% rms, and the width $0.034\pm0.006$ Hz.
 The shape of the power spectrum 
is typical for black hole candidates:
almost flat at frequencies lower than 0.03 Hz, 
roughly power law with slope $\sim$1.2 from 0.03 to 1 Hz,
with a following steepening to $\sim$1.6 at higher frequencies. 
The hard energy spectrum detected up to $\sim$150 keV
and the absence of significant X-ray variability at the high 
frequencies above 100 Hz strongly support the identification of
XTE J1118+480 as black hole transient.


\end{abstract}

\section{Introduction}
The transient X-ray source XTE J1118+480 was discovered 
with the RXTE All-Sky Monitor on March 29th, 2000. Subsequent RXTE pointed
observations revealed a power law energy spectrum 
with a photon index of about 1.8 up to at least 30 keV. 
No X-Ray pulsations were detected (Remillard et al., 2000)
In hard X-rays the source was
observed by BATSE up to 120 keV (Wilson\&McCollough
2000). 
Uemura, Kato \& Yamaoka(2000) reported  the optical counterpart of
12.9 magnitude in unfiltered CCD. The optical spectrum was found
typical for the spectrum of an X-Ray Nova in outburst
(Garcia et al. 2000). 
Pooley \& Waldram (2000) using Ryle Telescope detected a
noisy radiosource with flux density of 6.2 mJy at 15 GHz. 

All existing observations show that XTE J1118+480 
is similar to the black hole transients in close binaries 
with a low mass companion.

In this Letter we report on the detection of quasi-periodical 
oscillations (QPO) in the power spectrum of this source.

\section{Observations and data analysis}

\begin{figure}
\epsfxsize=8cm
\epsffile[30 180 530 720]{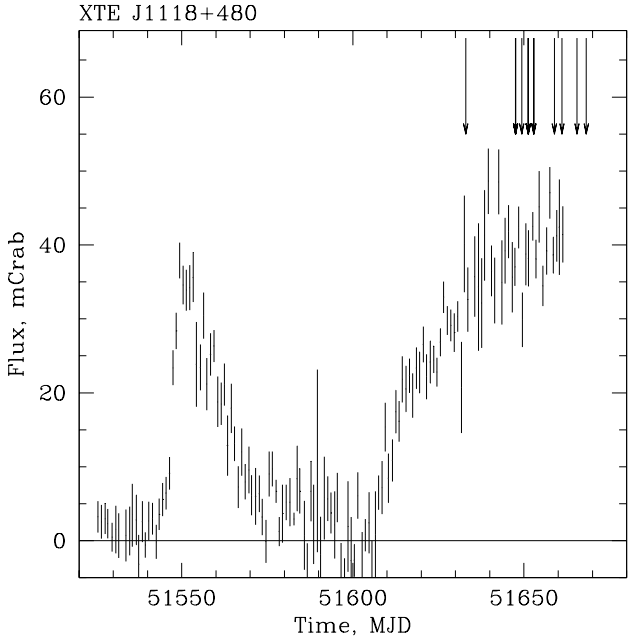}
\caption{The RXTE/ASM light curve (1.3-12.2 keV) of the
transient XTE J1118+480. Arrows show the dates of RXTE pointed
observations, used in our analysis.\label{lcurve}}
\end{figure}

In our analysis we used public domain target-of-opportunity
observations of XTE J1118+480 performed by the Rossi X-ray 
Timing Explorer (RXTE) observatory in the period Mar.29 -- May 4, 2000. 
The total usable exposure time of 11 observations was approximately 
22 ksec. The source was weak during all analyzed RXTE observations 
($\sim$40 mCrab in 2-12 keV spectral band).
In Fig. \ref{lcurve} we present a light curve of the recent outburst
of XTE J1118+480 accoring to data of All Sky Monitor (ASM) aboard RXTE as it is
provided by MIT ASM/RXTE team.

All data reduction was performed using the latest HEAsoft/FTOOLS 5.0
package. The correction for counting statistics noise in the power
spectra was done according to the procedure described in Sunyaev \&
Revnivtsev 2000 (see also \cite{zhangdt96}). 

\begin{table}
\caption{RXTE target-of-opportunity (TOO) observations of XTE
J1118+480 used in our analysis. \label{log}}
\tabcolsep=0.4cm
\begin{tabular}{ccccc}
\hline
\hline
ObsID&Date & TimeStart&Exp.\\
   &  2000    &   UT& ksec\\
\hline
50503-01-01-00&Mar.29&22:51&0.7\\
50407-01-01-00&Apr.13&09:28&5.0\\
50407-01-01-01&Apr.13&14:18&3.1\\
50407-01-02-00&Apr.15&07:51&1.1\\
50407-01-02-01&Apr.17&04:44&4.1\\
50407-01-02-02&Apr.18&19:21&1.0\\
50407-01-02-03&Apr.18&21:27&1.8\\
50407-01-03-01&Apr.24&20:35&0.7\\
50407-01-03-02&Apr.27&01:57&0.9\\
50407-01-04-02&May 1&11:25&1.8\\
50407-01-04-01&May 4&05:15&1.0\\
\hline
\end{tabular}

\end{table}

\section{Results}
\begin{figure}
\epsfxsize=8cm
\epsffile[30 180 530 720]{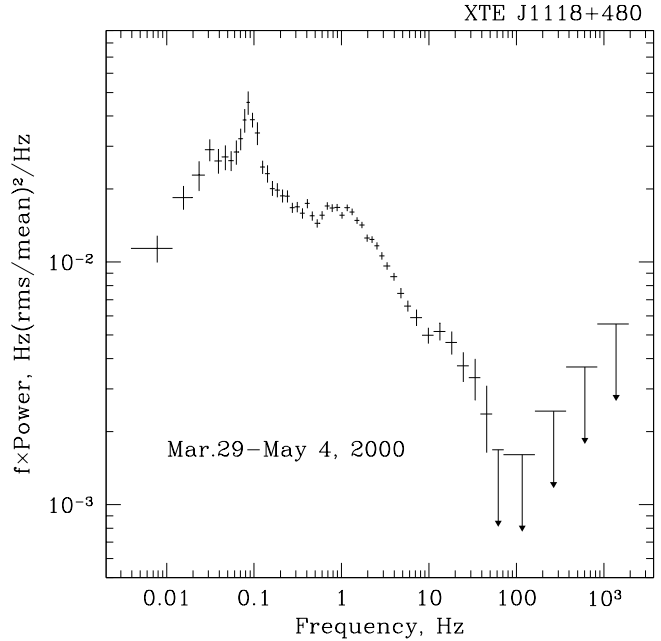}
\caption{Power spectrum of XTE J1118+480\label{power}}
\end{figure}

The power spectrum of XTE J1118+480 with a strong QPO feature 
is shown in Fig. \ref{power}. The simplest Lorenz approximation 
of the detected QPO peak gives the centroid frequency $0.085\pm0.002$ Hz 
and the width $0.034\pm0.006$ Hz (the $Q$ parameter $\sim$2--3). 
The amplitude of the QPO $\approx$10\% rms.
The power density spectrum (PDS) of the source is typical 
for a black hole candidates in the low/hard spectral state. 
The power spectrum is almost flat at frequencies below $\sim$0.03 Hz,
roughly a power law with slope $\sim$1.2 from 0.03 to 1 Hz with following 
steepening to slope $\sim$1.6 at higher frequencies. 
The total amplitude of detected variability of the source is close 
to 40\% rms. We did not detect any
X-ray variability of the source flux at the frequencies higher than
$\sim$70 Hz. The $2\sigma$ upper limits on the kHz QPOs in the
frequency band 300--1000 Hz are of the order of 5--6\% for QPO with
quality $Q\sim10$, this is in 1.5--3 times lower than typical amplitudes
of observed kHz QPOs in the neutron star PDSs (e.g. \cite{vdk}).

Our preliminary analysis of the XTE J1118+480 radiation spectrum confirms
that it is very hard: it was detected by High Energy 
Timing Experiment (HEXTE) up to energies of
$\sim$130--150 keV with the power law slope $\alpha\sim1.8$ with
possible cutoff at the highest energies ($\ga$130 keV)
The spectrum of XTE J1118+480
is very similar to that of the transient source GRS 1737--37
(\cite{disc1737}, \cite{tsp1737}, \cite{cui1737}). A detailed spectral analysis 
of XTE J1118+480 will be presented elsewhere.

\section{Discussion}
Low frequency QPO peaks were reported earlier in the power spectra 
of several black hole candidates in their low/hard state -- at 
$\sim$0.03--0.07 Hz with $Q\sim1$  for Cyg X-1 (Vikhlinin et al. 
1992, 1994,  Kouveliotou et al. 1992a), 
at $\sim$0.3 Hz for GRO J0422+32 (Kouveliotou et
al. 1992b, Vikhlinin et al. 1995), $\sim$0.8 Hz for GX 339-4 
(e.g. Grebenev et al, 1991) and in the high/soft state of LMC X-1
(\cite{eb_lmcx1}) and XTE J1748--288 (\cite{we_1748}). Impressive QPOs
with harmonics   
were observed in the power spectra of Nova Muscae 1991
(e.g. \cite{jap_muscae}, \cite{bel_qpo}), GRS 1915+105 
(e.g. Greiner et al. 1996, Trudolyubov et al. 1999b). 
The detection of low frequency QPO in the power spectrum of
XTE J1118+480 allows us to add another black hole candidate to this
sample.  In all these cases the QPO
peak lies close to the first (low frequency) break in 
the power spectrum (see also \cite{wvdk}).

The optical counterpart of XTE J1118+480 is sufficiently bright 
to check for the presence of corresponding low frequency 
optical variability with $f\sim0.085$ Hz.

The power spectra of black hole candidates are drastically 
different from those of neutron stars in LMXBs in similar 
low/hard spectral state. Sunyaev and Revnivtsev (2000) 
presented a comparison of power spectra for 9 black
hole candidates and 9 neutron stars. None of the black hole candidates
from this sample show a significant variability 
above $\sim$100 Hz, while all 9 neutron stars were noisy 
well above 500 Hz, with the significant contribution of 
high-frequency noise $f>150$ Hz to the total variability of the source. 
The power spectrum of the newly discovered X-ray transient 
XTE J1118+480 (see Fig 2) looks very similar to other 
black hole PDSs (see Fig.1 of Sunyaev and Revnivtsev, 2000).

The detection of low frequency QPO, lack of high-frequency noise 
and a hard energy spectrum detected up to $\sim$150 keV in X-rays
are supportive arguments for the earlier identification of
XTE J1118+480 as a black hole candidate.

\begin{acknowledgements}
This research has made use of data obtained 
through the High Energy Astrophysics Science Archive Research Center
Online Service, provided by the NASA/Goddard Space Flight Center.
The work has been supported in part by RFBR grant 00-15-96649.
\end{acknowledgements}

\end{document}